\documentclass[preprint]{aastex}
\usepackage{threeparttable}
\shorttitle{The central engine of 3C120}
\shortauthors{Cowperthwaite \and Reynolds}

\begin{document}


\title{The Central Engine Structure of 3C120: Evidence for a Retrograde Black Hole or a Refilling Accretion Disk}


\author{Philip S. Cowperthwaite \and Christopher S. Reynolds} 
\affil{Department of Astronomy, University of Maryland, College Park, MD 20742-2421, USA}
\affil{Joint Space-Science Institute (JSI), College Park, MD 20742-2421, USA}


%\altaffiltext{1}{Department of Astronomy, University of Maryland, College Park, MD 20742-2421, USA; pcowpert@umd.edu}
%\altaffiltext{2}{Joint Space-Science Institute (JSI), College Park, MD 20742-2421, USA}


\begin{abstract}
The broad-line radio galaxy 3C120 is a powerful source of both X-ray and radio emission including superluminal jet outflows. We report on our reanalysis of 160\,ks of {\it Suzaku} data taken in 2006, previously examined by Kataoka et al. (2007). Spectral fits to the XIS and HXD/PIN data over a range of 0.7--45\,keV reveal a well-defined iron K line complex with a narrow K$\alpha$ core and relativistically broadened features consistent with emission from the inner regions of the accretion disk. Furthermore, the inner region of the disk appears to be truncated with an inner radius of $r_{in} = 11.7^{+3.5}_{-5.2}\,r_g$. If we assume that fluorescent iron line features terminate at the inner-most stable circular orbit (ISCO), we measure a black hole spin of $\hat{a} < -0.1$ at a 90\% confidence level. A rapidly spinning prograde black hole ($\hat{a} > 0.8$) can be ruled out at the 99\% confidence level. Alternatively, the disk may be truncated well outside of the ISCO of a rapid prograde hole. The most compelling scenario is the possibility that the inner regions of the disk were destroyed/ejected by catastrophic instabilities just prior to the time these observations were made. 
\end{abstract}



\section{Introduction}

The central engines of active galactic nuclei (AGN) are powered by the accretion of material onto a supermassive black hole (SMBH). Some fraction of the AGN population, the radio-loud AGN, produce powerful, relativistic jets (Donoso, Best, \& Kauffmann 2009) --- but the physical processes that distinguish radio-loud from radio-quiet AGN remain mysterious.   It has been suggested that the spin of the SMBH plays an important role in determining whether an AGN is radio-loud or radio-quiet, a hypothesis known as the ``Spin Paradigm" (Sikora et al. 2007).   It is commonly accepted that relativistic jets are powered by the extraction of rotational energy from the ergoregion of a spinning black hole (Blandford \& Znajek 1977; McKinney, Tchekhovskoy, \& Blandford 2012).   However, whether the spin determines the radio-loudness of an AGN (i.e. whether there is an additional ``clutch" that switches on/off the transmission of spin power into the jet) is still very much up for debate.

The spin of a black hole with angular momentum $J$ is described by the dimensionless spin-parameter $\hat{a} = cJ / GM^2$ where $|\hat{a}| \le 1$. In order to test the spin paradigm it is important that we measure the spin parameter in both radio-quiet and radio-loud sources. Previous measurements of several radio-quiet objects have found the central SMBH to be in a state of rapid prograde spin (Brenneman et al. 2011); this already presents a challenge to the spin paradigm. Our goal is to address the other side of this problem by studying the central engine structure and black hole spin of a radio-loud object.  

The current best technique for measuring SMBH spin is by studying the relativistically broadened spectral features produced when the surface of the accretion disk is irradiated by hard X-rays. The most prominent of these reflection spectrum features is the fluorescent {\sc Fe}-K$\alpha$ emission line found at 6.4\,keV in the rest frame. The line emanates from the inner most regions of the accretion disk where it is broadened and skewed towards low energies by gravitational redshift and relativistic Doppler shift (Fabian et al. 1989). The line is clearly visible against the background continuum spectrum and serves as the strongest diagnostic for gravitational physics around SMBHs currently known (see reviews by Reynolds \& Nowak 2003; Miller 2007). If one assumes that the line emitting region of an accretion disk extends down to the inner-most stable circular orbit (ISCO) (Reynolds \& Fabian 2008) then, by modeling the profile of the broad {\sc Fe}-K$\alpha$ line, the location of the ISCO can be discerned in units of $r_g \equiv GM / c^2$, thereby giving us the spin (Brenneman \& Reynolds, 2006).

The broad-line radio galaxy (BLRG) 3C120 (z=0.033) is a prime candidate for such a study. It is the brightest BLRG in the X-ray sky and possesses a powerful one-sided jet on 100\,kpc scales (Walker et al. 1987, Harris et al. 2004) with observed superluminal motion ($\beta_{app} = 8.1$) indicating a jet inclination of $i < 14^{\circ}$ (Zensus 1989, Eracleous \& Halpern 1998).  Early X-ray observations by ASCA detected an extremely broad iron K$\alpha$ line (Grandi et al. 1997, Reynolds 1997, Sambruna et al. 1999). A significantly narrower line complex was found in later observations by RXTE (Eracleous, et al. 2000, Gliozzi et al. 2003) and BeppoSAX (Zdziarski \& Grandi 2001). Analyses of XMM-Newton data produced mixed results, with Ballantyne, Fabian \& Iwasawa (2004) finding a weak broad iron line (albeit with a poorly constrained inner emission radius) but Ogle et al. (2005) claiming no relativistic line signatures.  Aided by the hard X-ray coverage, observations of 3C120 by {\it Suzaku}    found both narrow iron emission lines and features consistent with a relativistically broadened line (Kataoka et al. 2007; hereafter K07). K07 were able to constrain the inner edge of the line emitting region to be $r_{\rm in}=8.6_{-0.6}^{+1.0}\,r_g$.   While not commented upon in K07, this begs the question as to whether 3C120 may be driven by a rapidly retrograde spinning SMBH (whose ISCO is at $9\,r_g$). 

In this \emph{Letter}, we present a reanalysis of the 2006 {\it Suzaku} data on 3C120 using updated instrument calibrations and the latest models/methodology for studying relativistically broadened reflection features.  Section 2 contains an outline of the observations and data reduction steps taken. Section 3 presents the spectral analysis which fundamentally reconfirms the results of K07, and Section 4 discusses the implications of our analysis.   Our results indicate either a retrograde spinning black hole, or a disk that is truncated beyond its ISCO.  We suggest a possible relation to the disk-jet cycles observed to occur in BLRGs such as 3C120.  

\section{Observations and Data Reduction}

The radio galaxy 3C120 was observed with {\it Suzaku} starting on 2006 February 7 beginning at 03:20 (UT) and ending on 2006 March 03 20:29 (UT). A total of four quasi-continuous observations were made, each with approximately 40\,ks of good on-source exposure. {\it Suzaku} is equipped with four X-ray CCD cameras, X-ray Imaging Spectrometer (XIS), which are sensitive from 0.3--12\,keV. Data were collected from all four XISs as these observations were made while XIS2 was still operational. Additionally, there is a Hard X-ray Detector (HXD) which is sensitive from 10--600\,keV. It consists of Positive Intrinsic Negative (PIN) silicon diodes and Gadolinium Silicate (GSO) phoswitch counters. All four observations were made with 3C120 placed at the nominal aim point for the XIS detector. 

Data reduction was performed using the HEASOFT data reduction suite. The event files were cleaned using standard filtering criteria and spectra were extracted with the {\tt xselect} software as per the {\it Suzaku} ABC guide. The most recent CALDB files at the time of analysis (Fall 2011) were used. The 3x3 and 5x5 event files were combined within {\tt xselect} prior to spectral extraction. The cleaned event files contained 144\,ks of good data. Spectra were extracted from counts within a 208" circular region centered on the source. Background spectra were produced from an annular region around the source. We generated response files for each XIS sensor and each pointing using the {\tt xisrmfgen} and {\tt xissimarfgen} scripts. Lastly, for each XIS, the 4 pointings were co-added; however we did not co-add spectra from the different XISs. This allows us to remain sensitive to cross-calibration issues.

Similarly, the HXD data was reduced as per the ABC guide. The HXD/PIN was able to detect 3C120 while the HXD/GSO was not. We produced cleaned event files and extracted spectra using the standard tools. These were then dead time corrected using the {\tt hxddtcor} script. We obtained the appropriate ``tuned" non-X-ray background files (NXB) for these observations. Epoch appropriate response files were obtained from the {\it Suzaku} CALDB website. The Cosmic X-ray Background (CXB) files were simulated with XSPEC 12.7 using the provided response file and following the typical model based on HEAO1 results (Boldt 1987). The CXB and NXB files were then added to produce a final background spectra. As with the XIS data, the four HXD/PIN observations and their backgrounds were co-added.

\section{Spectral Analysis}

All of the spectral analysis performed for this paper used XSPEC v12.7 (Arnaud 1996). Errors were computed using Monte Carlo Markov Chain (MCMC) methods, and are reported at the 90\% level of confidence. The fitting parameters for our spectral model were tied across the HXD/PIN and XIS detectors save for a cross-normalization factor to compensate for differences in the absolute flux between detectors. The seed value for the HXD/PIN cross-normalization was 1.16 in order to be consistent with current recommended calibrations ({\it Suzaku} Memo 2008-06) but, as we note below, the data strongly requires an HXD/PIN cross-normalization closer to unity.

We begin by examining the spectra in the ranges of 0.7--1.5\,keV and 2.5--10\,keV. The data in the 1.5--2.5\,keV band were ignored during all fits due to known issues with the XIS calibration over this range of energies. Our initial model consisted of a simple absorbed power-law. Figure 1 zooms in on the 3--9\,keV band and shows that this model results in a poor fit to the data ($\chi^2 / \nu= 6556/4768\;[1.37]$) but illuminates a variety of interesting spectral features. The most noticeable of such features is the strong emission line at 6.4\,keV in the source rest frame which is likely due to the {\sc Fe}-K$\alpha$ fluorescence line.  This line has a red wing which extends to $\sim$5\,keV indicating possible relativistic broadening effects. There is also a weaker emission feature at $\sim$6.9\,keV which is likely a blend of the cold {\sc Fe}-K$\beta$ line and the {\sc Fe\,xxvi} line. The hard excess flux at energies greater than 8\,keV is interpreted as the Compton hump associated with the X-ray reflection producing these emission features.

Guided by this preliminary fit and taking guidance from K07, we built a heuristic model consisting of photoelectric absorption ({\tt phabs}), the continuum from reflection off cold matter ({\tt pexmon}, Nandra et al. 2007), a narrow Gaussian {\sc Fe\,xxvi}-K$\alpha$ line at 6.97\,keV, and an ionized reflection model ({\tt reflionx}, Ross \& Fabian 2005) convolved with a relativistic smearing model ({\tt relconv}, Dauser et al. 2010) to produce a smeared reflection profile for an ionized accretion disk. The {\tt pexmon} model is a particularly useful evolution of the {\tt pexrav} model of Magdziarz \& Zdziarski (1995). In addition to producing the Compton backscattered continuum it also generates physically self-consistent {\sc Fe}-K and {\sc Ni}-K lines.

Parameters common to {\tt pexmon} and  {\tt reflionx} were tied. Additionally, {\tt pexmon} takes the low-Z metal abundance and iron abundance ($Z_{\rm Fe}$) relative to solar values as fitting parameters. We allowed $Z_{\rm Fe}$ to be a free parameter, but the low-Z abundance was frozen at unity. The radial dependence of the emissivity of the disk reflection is assumed to be a power-law, $\epsilon\sim r^{-\beta}$; broken power-law profiles did not lead to a significant increase in the goodness of fit. Previous fits by K07 and Ogle et al. (2005) have assumed an emissivity index of $\beta = 3$, whereas we have allowed this value to be free.

We fit this model to the global XIS and HXD/PIN spectra over 0.7--45\,keV, considering two independent scenarios. Both scenarios are guided by the observation that radio flares are associated with X-ray dips (not flares), so we assume X-ray emission is dominated by the disk corona (Ogle et. al 2005, Marshall et al. 2009) and not the base of a jet. Scenario \#1 involves making the assumption that reflection components terminate at the ISCO and fitting for $\hat{a}$. We note the nominal range of spin values in {\tt relconv} is likely not appropriate, with MHD torques from the accretion flow limiting it to $|\hat{a}|<0.95$ (Penna et al. 2010). Thus, we have restricted the spin parameters that we consider to the range  $\hat{a}$ $\in [-0.95, 0.95]$. This produced a good fit to the data ($\chi^2 / \nu = 5527/5033 \;[1.10]$). We obtained a SMBH spin of $\hat{a} \le -0.1$, indicating a moderate to rapidly spinning retrograde hole (Table 1).  We also note a favorable result for the inclination of $\mathit{i} = 17.7^{+6.5}_{-8.7}$ degrees. This is in good agreement with the constraints derived from superluminal jet motion ($\mathit{i} < 14^{\circ}$) and suggests that the inner disk and jet are aligned as expected from the theory of Bardeen \& Petterson (1975).

Scenario \#2 investigates the possibility that the inner region of the disk is truncated beyond the ISCO, an approach similar to the original analysis of K07. With our model this is accomplished by freezing the spin at $\hat{a} = 0.95$ and allowing the inner disk radius to be a free parameter. The choice of a rapid prograde spin permits the model to integrate over the maximum range of allowable radii. The fit was nearly statistically identical to the ``spin-free" fit ($\chi^2 / \nu = 5524/5033 \;[1.10]$). The fitted inner radii was $r_{in} = 11.7^{+3.5}_{-5.2}\,r_g$ (Table 1). Essential disk parameters were \emph{not} preferentially seeded with results from the ``spin-free" fit; yet they still arrived at very similar values. The full set of parameters from both fits are listed in Table 1 (also see Figure 2). 

The strength of the narrow iron line was measured with a separate model consisting of an absorbed power law and two gaussian components at 6.4\,keV and 6.97\,keV. This allows us to isolate the narrow {\sc Fe}-K$\alpha$ line. The equivalent width was found to be EW$_{K\alpha}$ $= 55\pm6$\,eV. Following a similar prescription we isolated and measured the strength of the relativistically broadened line finding EW$_{Broad}$ $= 48\pm10$\,eV. We also computed model fluxes and luminosities which can be found in Table 2.

We note that a curious aspect in both of these scenarios is the behavior of the cross-normalization factor between the XIS and HXD/PIN datasets. The fiducial value put forth by the {\it Suzaku} science team, for this dataset, is $\tilde{C} = 1.16$. Both scenarios achieved a best fit with $C \sim 1$ which is not in statistical agreement with the fiducial value. In order to test the validity of this result we repeated our analysis, this time leaving the cross-normalization frozen at $\tilde{C}$. This produced a statistically worse fit to the data, in both scenarios, with $\Delta \chi^2 \sim 90$ for one additional degree of freedom. There are strong residuals above 8 keV which indicate an offset normalization for the HXD/PIN. We conclude that our model accurately describes the data and the fiducial cross-normalization factor is not applicable for this dataset. 

\section{Discussion \& Conclusion}

We confirm the presence of broad asymmetrical iron line features in 3C120 due to emission from the inner regions of an accretion disk. Interpreting the structure of the inner disk from these data we arrive at two possible conclusions. Under the assumption that fluorescent iron line emission terminates at the ISCO, our analysis of {\it Suzaku} data for 3C120 indicates that the central SMBH may be in a state of retrograde spin. The probability distribution for the spin parameter, derived from MCMC methods, confirms a spin of $a < -0.1$ at a 90\% confidence.  We can reject a rapidly rotating prograde black hole with spin $a > 0.8$ at the 99\% confidence level. We also consider evidence for a natural truncation of the disk, independent of the spin configuration of the central SMBH. In doing so we find that an inner disk radius of $r_{in} = 11.7^{+3.5}_{-5.2}\,r_g$ is strongly preferred by our model. Both probability distributions can be seen in Figure 3.

Theoretically, the possibility of a retrograde spin in 3C120 is intriguing.  Employing the flux trapping picture of Reynolds, Garofalo, \& Begelman (2006), Garofalo (2010) suggested that a rapidly-rotating retrograde black hole may be an effective configuration for producing a powerful jet, with the enlarged ISCO of the retrograde hole trapping a large bundle of magnetic flux and driving a powerful Blandford-Znajek jet.  Interestingly, this idea is not borne out in General Relativistic magnetohydrodynamic simulations which find that jets from prograde black holes are factors of several more powerful than from the equivalent retrograde hole (Tchekhovskoy et al. 2012; Tchekhovskoy \& McKinney 2012).   It is not clear, however, that the ``magnetically choked" configuration employed by Tchekhovskoy et al. is equivalent to the advective-diffusive equilibrium underlying the Garofalo (2010) models and so it is too early to say that such models have been invalidated by the simulations.  More theoretical work is required to assess the efficiency with which retrograde black holes can power jets.

It is equally intriguing to consider the implications of a natural truncation of the inner disk beyond the ISCO. One possibility is the disk undergoes a transition to a radiatively inefficient accretion flow (RIAF) at $r \sim 10\,r_g$.   This would cause the inner regions of the accretion disk to become optically thin and extremely hot, truncating iron line emission and all other signs of reflection. Such a transition is believed to occur once the accretion rate falls below an Eddington ratio ($\mathcal{L}$) of a few percent (Esin et al. 1997). For example, disk truncation is clearly measured in MCG -5-23-16 ($\mathcal{L}  = 0.05$; Reeves et al. 2006)  and NGC 4593 ($\mathcal{L}  = 0.04$; Markowitz \& Reeves 2009). However, 3C120 is not a low-$\mathcal{L}$ source. Using our model 2--10\,keV luminosity along with a bolometric correction of $\kappa_{2-10\,keV} = 20.6$ (Vasudevan \& Fabian 2009) and SMBH mass $M_{3C120} = 5.5 \times 10^7\,M_{\odot}$ (Peterson et al. 2004) we compute an Eddington ratio of $\mathcal{L}_{3C120} = 0.23$. This is high enough that a standard transition to a RIAF at $10r_g$ becomes unlikely.

A more intriguing possibility is that we are seeing a {\it transitory} truncation of the disk.  Long-term monitoring of 3C120 reveals periodic dips in X-ray luminosity coincident with (or slightly preceding) the ejection of a new superluminal VLBA knot in the jet accompanied by large radio bursts. It has been hypothesized that this may be due to a catastrophic instability, thermal or dynamic in nature, which completely destroys/ejects the inner regions of the disk; and that this process is an important component of jet formation (Marscher et al. 2002, Chatterjee et al. 2009, King et al. 2011). The destruction of the inner line emitting regions of the disk would cause iron line emission to cease.

Is it possible that such a dramatic event in the accretion disk of 3C120 occurred prior to the observations by {\it Suzaku}? In fact, a significant X-ray dip followed by a jet ejection event did indeed occur at the very beginning of 2006 ($2006.00\pm0.03$; Chatterjee et al. 2009), just one month prior to these {\it Suzaku} observations.  Following the destruction of the inner accretion disk, the inner regions need some amount of time to refill with material; so the issue is whether or not the disk had time to reform between the ejection event and the {\it Suzaku} observations.   At this point, it is instructive to consider relevant accretion disk timescales, focusing on a radius of $10\,r_g$ while assuming an angular momentum transport parameter of $\alpha = 0.1$ and  radiative efficiency of $\eta = 0.1$ for the disk.  The dynamical timescale is $t_{dyn}\equiv (r^3/GM)^{1/2}\approx 9ks$ and the viscous timescale is $t_{visc} = t_{dyn} / (\alpha [h/r]^2)$ where $h/r$ is the scale height of the disk. Accretion disk theory (Shakura \& Sunyaev 1973) states that the inner region of this radiation-pressure dominated accretion disk has a geometric thickness of $h = (3 \mathcal{L} / 2\eta)\;r_g.$ Using our computed Eddington ratio we find $h \approx 3.45 \;r_g$; which results in a viscous timescale of $t_{visc} \approx 750$\,ks (8.7 days)  at $10\,r_g$.  Given that the viscous timescale has a strong dependence on the precise choice of radius $(t_{visc} \propto r^{7/2}),$ it is reasonable to assume that this timescale may be easily extended to a month or more over the range of $r_{in}$ allowed by our fit.

We have presented a reanalysis of the 2006 {\it Suzaku} data for the AGN 3C120. Thanks to updated calibrations and models we were able to reconfirm and expand on prior analysis of this dataset by K07. We confirmed the presence of broad iron line features consistent with relativistically-blurred ionized reflection from the disk, and show that a disk truncation radius of $r\sim$ 7--14\,$r_g$ is preferred. We arrive at two possible explanations for the truncation; either this black hole is in a state of retrograde accretion, or we are seeing the refilling of the accretion disk following a jet-ejection event.   Detailed spectral monitoring of BLRGs with VLBA in conjunction with XMM-Newton or {\it Suzaku} are required to assess this last possibility --- this should be a key goal of future programs.

We thank Andy Fabian, Ashley King, Anne Lohfink, Cole Miller, and Margaret Trippe for stimulating discussions during the course of this work. We also thank our anonymous referee whose comments greatly improved this manuscript. This research has made use of data obtained from the High Energy Astrophysics Science Archive Research Center (HEASARC), provided by NASA's Goddard Space Flight Center.  CSR acknowledges support from NASA through the Suzaku Cycle-5 GO Program (grant NNX10AR31G) and the ADAP (grant NNX12AE13G).

\clearpage
\begin{figure}
\epsscale{0.9}
\plotone{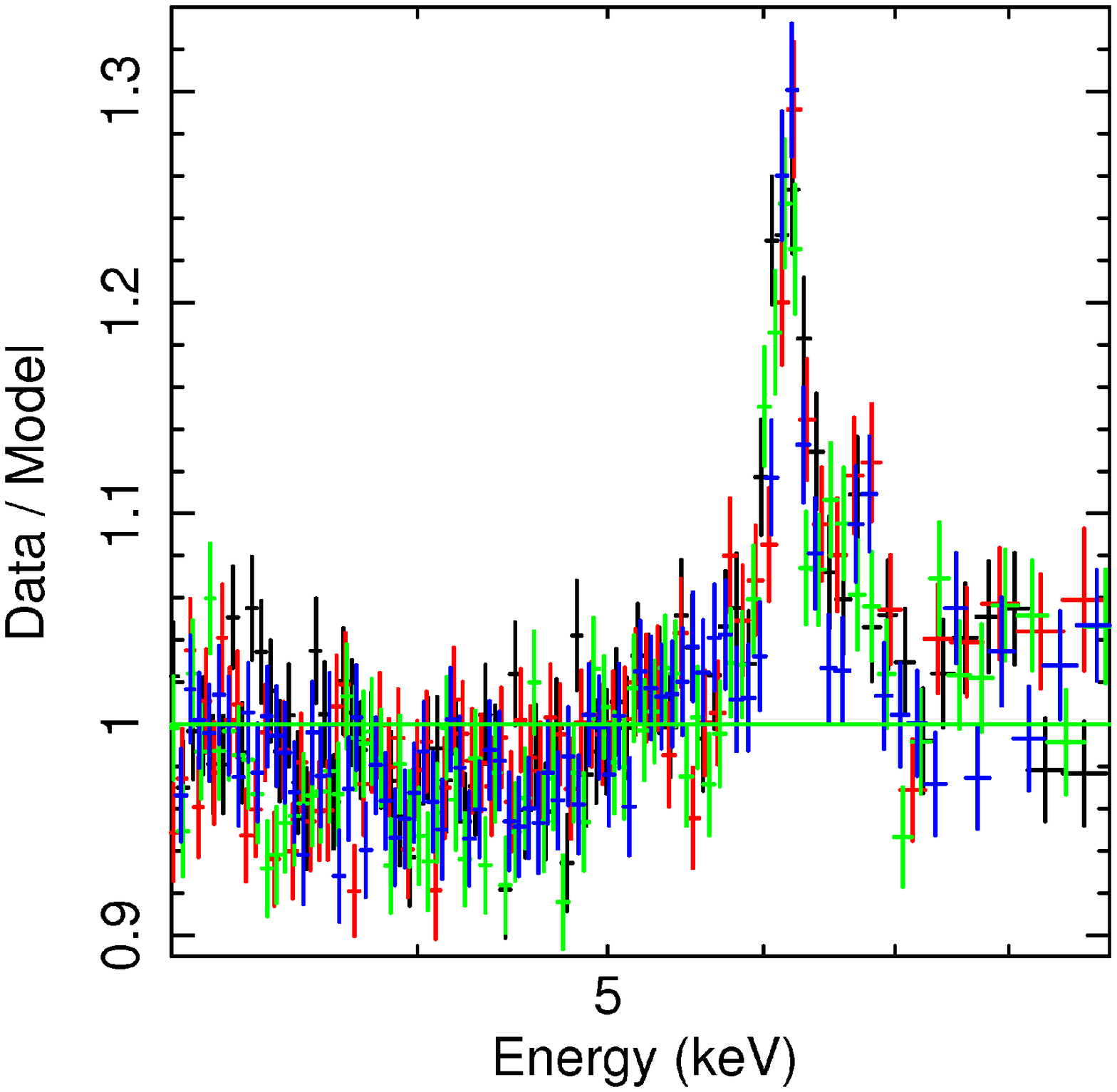}
\caption{The data/model residuals of the simple absorbed power-law fit to the XIS data in the 3--9\,keV range. The iron line complex is clearly visible around 6.4\,keV.}
\end{figure}

\begin{figure}
\epsscale{0.9}
\plotone{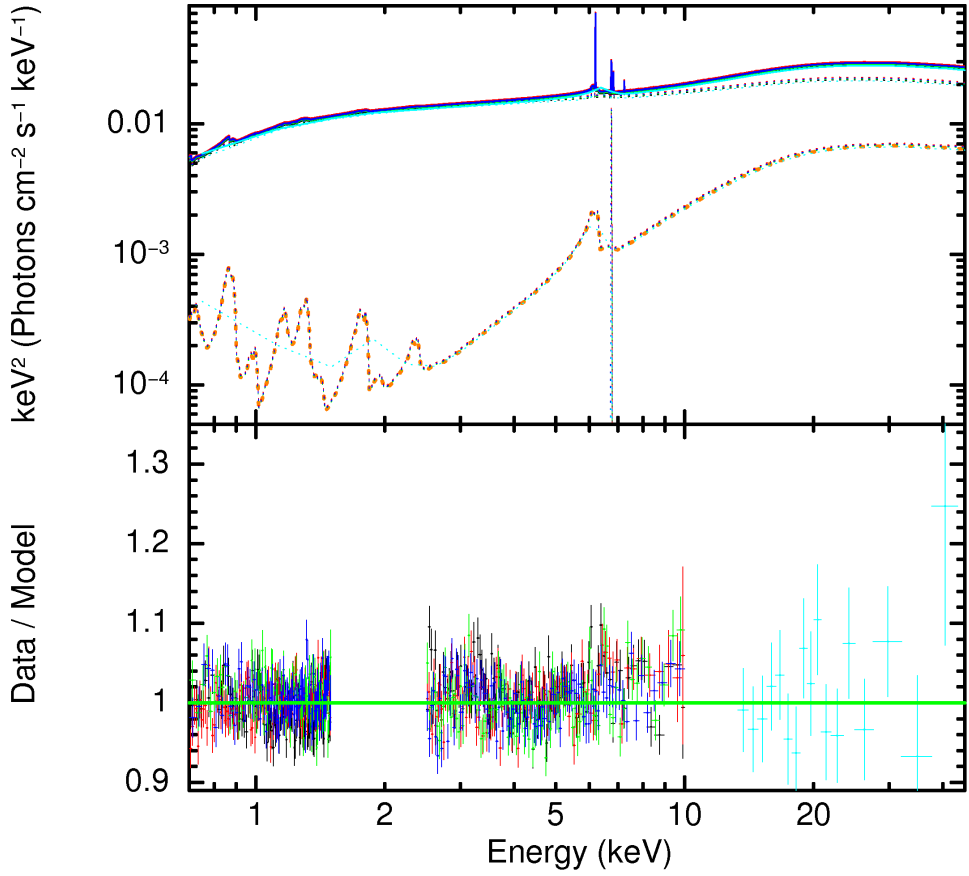}
\caption{The best-fit model to the global XIS+HXD/PIN spectrum from 0.7--45\,keV. The top panel shows the individual model components, multiplied by two factors of energy. The bottom panel shows the data/model residuals for the ``spin-free" fit. The residuals for the ``$r_{in}$-free fit" are nearly identical.}
\end{figure}

\begin{figure}
\epsscale{1.1}
\plottwo{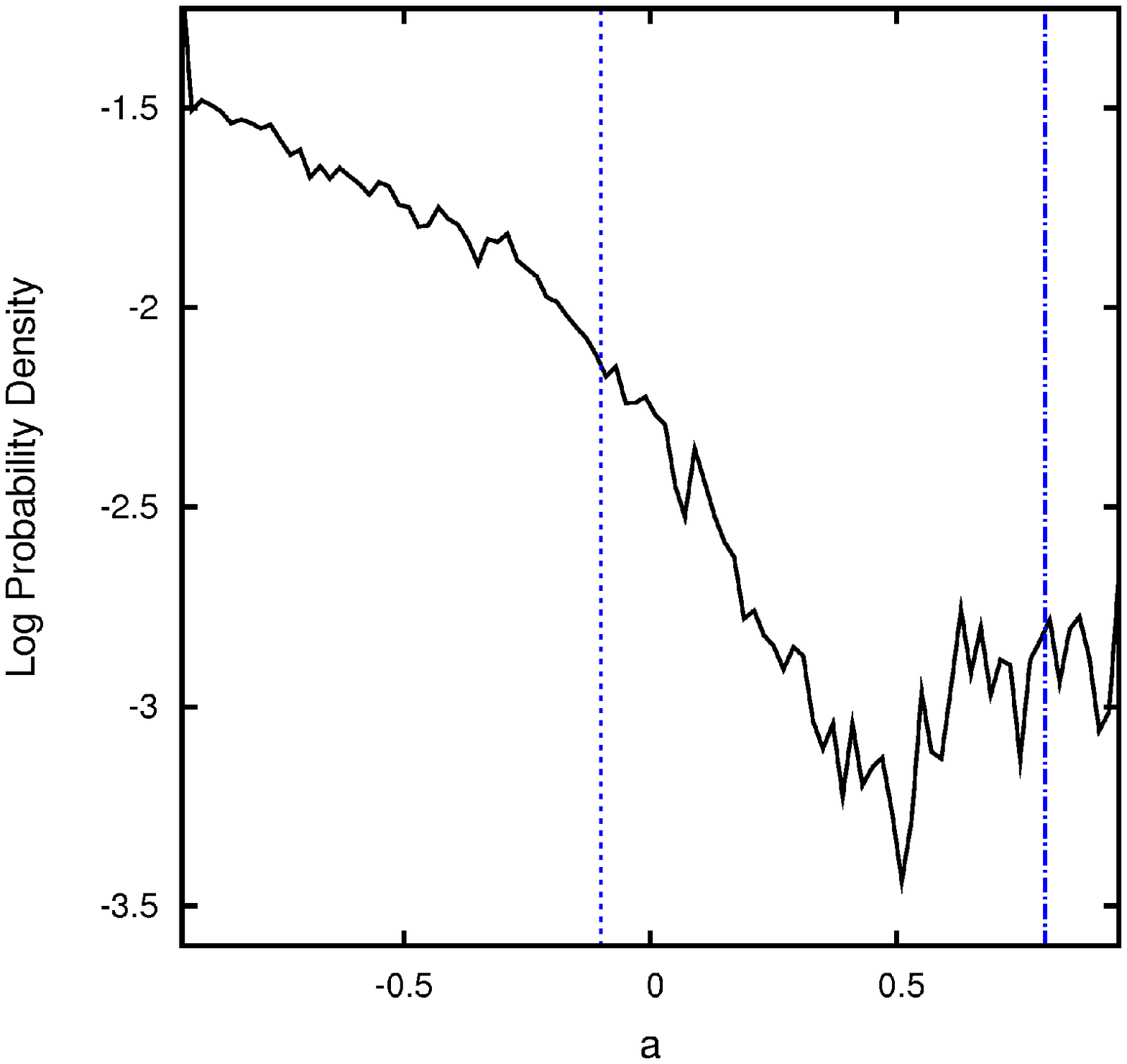}{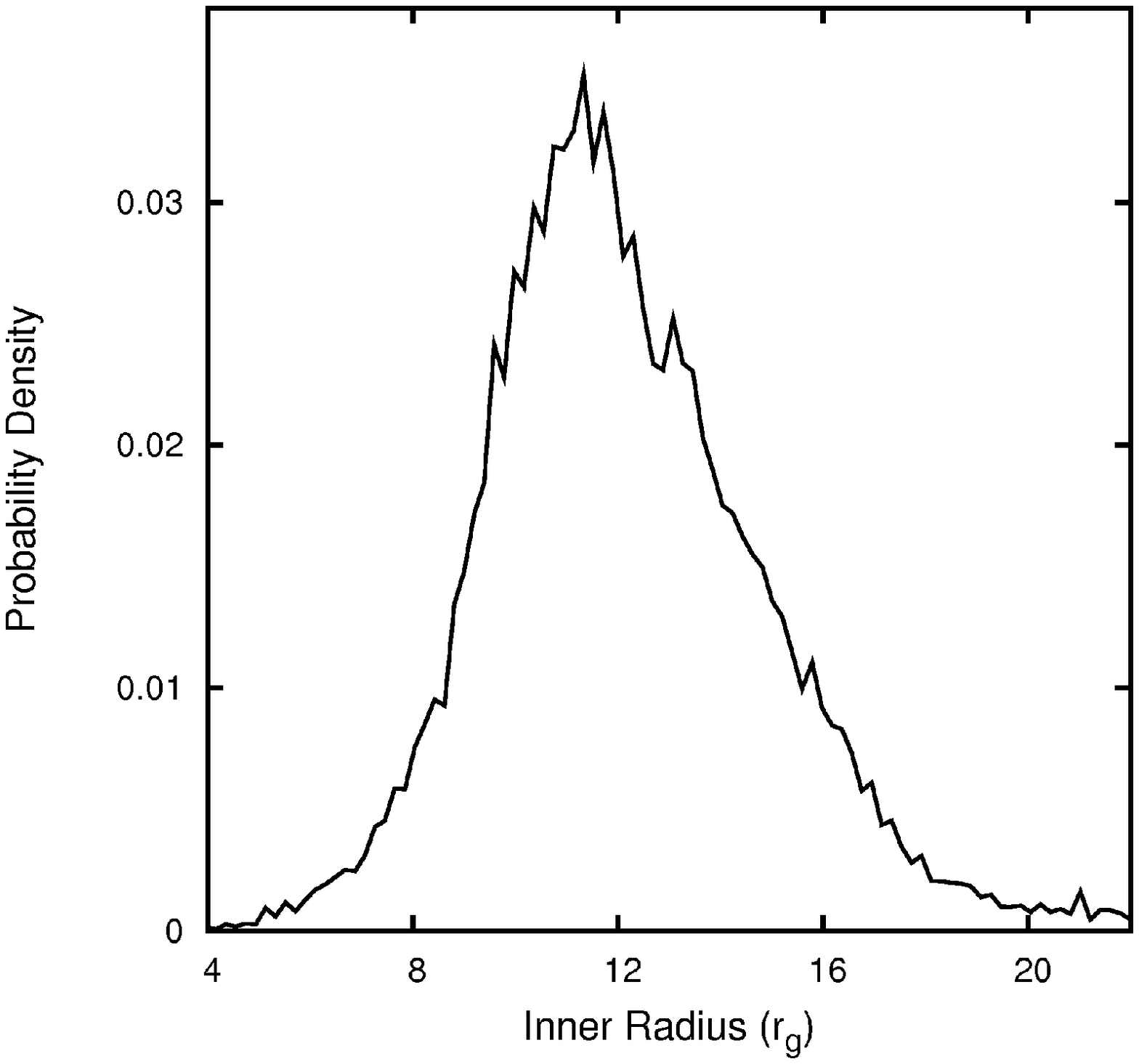}
\caption{The probability distributions for $\hat{a}$ and $r_{in}$. The distribution for spin was strongly peaked for negative values of $\hat{a}$. The choice of plotting $\log_{10} P(\hat{a})$ allows the structure of the distribution to remain discernible. The dashed and dot-dashed lines indicate the 90\% and 99\% CL upper limits, respectively.}
\end{figure}

\clearpage
\begin{table}
\begin{center}
\caption{Spectral Fitting Parameters}
\begin{tabular}{ c  c  c  c}
\\
\hline \hline
Model Component & Parameter & {\it Suzaku} (0.7--45\,keV) & {\it Suzaku} (0.7--45\,keV) \\
\hline
Galactic Column & N$_H$ & $0.158^{+0.001}_{-0.001}$ & $0.158^{+0.001}_{-0.001}$ \\
\hline
PL & $\Gamma$ & $1.87^{+0.01}_{-0.01}$ & $1.86^{+0.01}_{-0.01}$ \\
 & A$_{PL}$ & $(1.24^{+0.01}_{-0.01}) \times 10^{-2}$ & $(1.25^{+0.01}_{-0.01}) \times 10^{-2}$ \\
 \hline
 Cold Reflection & $\mathcal{R}$ & $0.56^{+0.05}_{-0.04}$ & $0.57^{+0.04}_{-0.05}$ \\
  & PL Cutoff (keV) & 150* & 150* \\
  \hline
  {\sc Fe\,xxvi}-K$\alpha$ Line & E (keV) & 6.97* & 6.97* \\
  & $\sigma$ (keV) & 0.01* & 0.01*  \\
  & A$_{Line}$ & $(7.0^{+1.5}_{-2.6}) \times 10^{-6}$ & $(6.9^{+1.5}_{-1.7}) \times 10^{-6}$ \\
  \hline
  Accretion Disk & Z$_{\rm Fe}$ & $0.50^{+0.06}_{-0.03}$ & $0.50^{+0.06}_{-0.04}$ \\
  & $\xi$ & $20^{+4}_{-3}$ & $21^{+3}_{-5}$   \\
  & $\mathit{i}$ &$17.7^{+6.5}_{-8.7}$ &$18.2^{+5.2}_{-0.9}$  \\
  & $\beta$ &  $2.38^{+0.23}_{-0.30}$ &  $2.63^{+0.69}_{-0.36}$ \\
  & r$_{in}$ & ISCO* &  $11.7^{+3.5}_{-5.2}$  \\
  & r$_{out}$ & 400* & 400*\\
  \hline
  PIN/XIS Normalization &  & $0.98^{+0.02}_{-0.03}$ & $0.98^{+0.02}_{-0.02}$\\
  \hline
  SMBH Spin & $\hat{a}$ &  $< -0.1$ &  0.95*\\
  \hline
  $\chi^2 / \nu$ & & $5527/5033 \;[1.10]$ & $5524/5033 \;[1.10]$\\
   \hline \hline
   \end{tabular}
   \end{center}
   \begin{tablenotes}
          \item Note --- All errors are quoted at the 90\% confidence level for one interesting parameter ($\Delta \chi^2 = 2.7$). Parameters which were fixed for the fit are marked with an asterisk. The units are as follows: column density is in $10^{22}$\,cm$^{-2}$, normalizations are in photons cm$^{-2}$\,s$^{-1}$, iron abundance is relative to solar values and linked between {\tt pexmon} and {\tt reflionx}, inclination is in degrees, all radii are in $r_g$, and spin is dimensionless (See text for details).
      \end{tablenotes}
      \vspace{5pt}
\end{table}

\clearpage
\begin{table}
\begin{center}
\caption {Model Fluxes and Luminosities for 3C120}
\vspace{-5pt}
\begin{tabular}{ c  c  c  c}
\\
\hline \hline
 & 0.7--2\,keV & 2--10\,keV & 0.7--10\,keV \\
\hline
Flux & $1.61\times10^{-11}$ & $4.03\times10^{-11}$ & $5.64\times10^{-11}$  \\
Flux$^{\dagger}$ & $2.20\times10^{-11}$ & $4.17\times10^{-11}$ & $6.39\times10^{-11}$  \\
Luminosity & $3.95\times10^{43}$ & $1.00\times10^{44}$ & $1.40\times10^{44}$ \\
Luminosity$^{\dagger}$ & $5.52\times10^{43}$& $1.05\times10^{44}$ & $1.60\times10^{44}$ \\
   \hline \hline
   \end{tabular}
   \end{center}
   \begin{tablenotes}
   \item Note --- The model fluxes and luminosities for 3C120. Fluxes are given in erg cm$^{-2}$ s$^{-1}$. Luminosities are computed with a standard cosmological model of $H_0 = 70$\,km s$^{-1}$ Mpc$^{-1}$, $\Omega_{\Lambda} = 0.73$ and $\Omega_{M} = 0.28$ and are given in erg s$^{-1}$. Values marked with a $\dagger$ have been absorption corrected. The errors are dominated by systematic effects and thus not reported.
   \end{tablenotes}
   \vspace{5pt}
\end{table}

%We have used our model to compute the fluxes and luminosities for the source. The observed frame fluxes are $F_{2-10\,keV} = 4.03\times10^{-11}$\,erg cm$^{-2}$ s$^{-1}$ and $F_{0.7-10\,keV} = 5.61\times10^{-11}$\,erg cm$^{-2}$ s$^{-1}$. Implementing a cosmological model of  we computed source frame luminosities of $L_{2-10\,keV} = 1.00\times10^{44}$\,erg s$^{-1}$ and $L_{0.7-10\,keV} = 1.39\times10^{44}$\,erg s$^{-1}$.

\end{document}